\documentclass[doublecol]{epl2} 
\usepackage{graphicx}
\usepackage{amssymb,amsmath,amsfonts}
\usepackage{array}
\usepackage{dcolumn}
\usepackage{color}

\title{Accurate evaluation of magnetic coupling between atoms with numerous
  open-shells~: an ab-initio method}

\author{Alain Gell\'e, Julien Varignon \and Marie-Bernadette Lepetit}

\institute{CRISMAT, ENSICAEN-CNRS UMR6508, 6~bd. Mar\'echal Juin,
14050 Caen, FRANCE}

\date{\today}

\abstract{
  We propose a new ab initio method designed for the accurate calculation of
  effective exchange integrals between atoms with numerous open-shells. This
  method applies to ferromagnetic as well as antiferromagnetic exchange,
  direct or ligand-mediated exchange. Test calculations on high spin
  transition metal oxides such as $\rm KNiF_3$, $\rm Ba_2CoS_3$ or $\rm
  YMnO_3$ exhibit a very good accuracy compared either to the best ab initio
  calculations ---~when those are feasible~--- and with experimental
  evaluations.
}


\begin{document}
\maketitle



In the last two decades transition metal oxides attracted a lot of attention
due to the discovery of novel and intriguing properties.  The interest of the
physicists community was first triggered by the discovery of high-temperature
superconductivity in copper oxide compounds. Since then, many transition metal
oxides exhibiting properties of practical or theoretical interest were
discovered. Presently the attention of the community is mainly focused on
cobalt, manganese and iron oxides. Most of these systems present intriguing
magnetic properties or ordering. 
It is thus of crucial importance for the understanding of these properties to
be able to accurately evaluate the local effective interactions such as
magnetic interactions between the Fermi-level, unpaired, electrons.

In these materials, the Fermi-level electrons are localized, essentially on
the $d$ orbitals of the metal ions. This localization is a consequence of the
fact that the electron-electron repulsion is of larger magnitude than the
kinetic energy. Such strongly correlated materials cannot be described by a
simple band structure since such a description supposes the dominance of the
delocalization effects over electron-electron repulsion ones.  The importance
of the electronic correlation results in charge, spin or orbitals occupation
fluctuations arising from the competition between different configurations
in the electronic structure. The nature of the ground and the low-lying
excited-states is thus fundamentally multi-configurational. Therefore it is not
surprising that ab-initio single-determinant based methods (such as density
functional theory) encounter difficulties in properly describing strongly
correlated systems. For this purpose physicists use model Hamiltonians. The
knowledge of the pertinent degrees of freedom to be treated in such models and
the amplitude of the interactions between them is thus of crucial importance.

In the case of copper, nickel or vanadium oxides, such microscopic models were
accurately determined using exact diagonalization of selected configurations
spaces (Configuration Interaction or CI methods) on embedded
fragments~\cite{Revue,env}. The CAS+DDCI~\cite{DDCI} (Complete Active Space +
Difference Dedicated Configurations Interaction also found under the
  acronym of DDCI3) and LCAS+S~\cite{lcas+s}
(Large CAS + single excitations) type of configurations selections proved
their high reliability and efficiency. Indeed, they allowed the determination
of magnetic couplings within experimental accuracy for the superconducting
copper oxides~\cite{CALZADO99,david}. In molecular systems, where the magnetic
coupling can be experimentally evaluated with a great accuracy, the cited
methods were able to reproduce them with an error smaller than $5\rm\
cm^{-1}$~\cite{molmagn}. Moreover the availability of the wave function offers
the possibility to evaluate the pertinence of the chosen effective model. Let
us recall the example of the famous $\rm \alpha^\prime NaV_2O_5$ where such ab
initio study showed the incompleteness of the commonly used models that were
neglecting the magnetic role of the oxygen ligands. The novel model, derived
using CAS+DDCI ab initio calculations, allowed to explain the apparently
contradictory experimental results of this compound~\cite{nav2o5}.

Unfortunately the CAS+DDCI, LCAS+S and related methods cannot be used for
systems involving more than one or two unpaired electrons per magnetic center,
and therefore not for high spin manganese, cobalt, iron oxides, etc. Indeed,
for such systems the computational cost of the methods becomes prohibitive
since the size of the space to diagonalize scales exponentially with the
number of magnetic electrons. In this paper we propose an ab initio approach
that overcomes this problem. We establish a simple physical criterion in order
to select the important reference configurations, and derive from it a novel
method with a strongly reduced computational cost. We will see that this
method allows, for the first time, the determination within experimental
accuracy, of effective magnetic interactions in high spin manganese or cobalt
oxides, thus opening entire new classes of materials to ab initio studies.


Let us first recall the principle of the usual CAS+DDCI and LCAS+S methods.
These variational methods are based on the expansion of the low lying states
wave functions into a reference part and a screening part. 
 \begin{equation}
  |\Psi_m\rangle  = \overbrace{\sum_I C_{I,m} |\Phi_I\rangle}^{\text{reference}} + 
 \overbrace{\sum_{J^*} C_{J^*,m} |\Phi_{J^*}\rangle}^{\text{screening}}
=  |\Psi_m^r\rangle +   |\Psi_m^*\rangle 
\end{equation}
In the CAS+DDCI and LCAS+S approaches the reference configurations,
$|\Phi_I\rangle$, span a Complete Active Space (CAS) and the screening
configurations, $|\Psi_m^*\rangle$, are all single (or single and selected
double) excitations on all the references $|\Phi_I\rangle$.  The definition
of the CAS is based on a mathematical formalization of the distinction between
the core or ligands electrons ---~that are essentially paired~--- and the
essentially unpaired magnetic electrons.  Indeed, the orbital space is
partitioned into three subsets~: i) the {\em occupied orbitals} that are
doubly occupied in all the CAS configurations, ii) the {\em active orbitals}
that can take any occupation or spin values in the CAS configurations and iii)
the {\em virtual orbitals} that are unoccupied in all CAS configurations. In
the CAS+DDCI method the three orbitals sets strictly correspond to the orbitals
supporting the essentially paired electrons, the magnetic electrons, and the
rest. The screening configurations are then taken as all the single and double
excitations on any of the CAS configurations, that contribute to the
excitation energies at the second order of perturbation.

In the LCAS+S method one uses the fact that the screening effects are
essentially supported by single excitations. The double excitations are thus
eliminated from the calculations under the condition that the configurations
associated with all important processes mediating the interactions between the
magnetic atoms are incorporated within the references~\cite{lcas+s}. The
LCAS+S method thus uses an active orbital set enlarged to the ligand orbitals
mediating the interactions between the magnetic orbitals. The screening part
is thus reduced to the single excitations on this large CAS.  It was shown
that the CAS+DDCI and LCAS+S methods yield results of equivalent
accuracy~\cite{lcas+s}, thus rooting the idea that the important effects to
treat, in order to achieve accurate evaluation of local interactions, are the
interactions between i) the magnetic configurations (referred to as
zeroth-order) ii) the charge transfer configurations mediating the
interactions (metal-to-metal and ligand-to-metal) and iii) the screening
effects on the first two subsets as given by the single excitations on them.
\begin{eqnarray} \label{lcas}
  |\Psi_m\rangle &=&  \overbrace{
    \underbrace{\sum_{I} C_{I,m}^0 |\Phi_I^0\rangle}_{\text{zeroth-order}} +  
    \underbrace{\sum_J C_{J,m} |\Phi_J\rangle}_{\text{charge transfer}}}^{\text{reference}} +
  \overbrace{\sum_{J^*} C_{J^*,m} |\Phi_{J^*}\rangle}^{\text{screening}}
  \nonumber \\
  &=&  |\Psi_m^0\rangle +  |\Psi_m^{ct}\rangle  + |\Psi_m^*\rangle
\end{eqnarray}

The size of the configurations space to diagonalize in both the CAS+DDCI and
LCAS+S methods is directly proportional to the size of the CAS that itself
scales exponentially with the number of active orbitals. Thus, when the number
of open shells per atom, or the number of magnetic atoms increases, it rapidly
results in intractable calculations. For instance, the evaluation of the
magnetic coupling between two $\rm Mn^{3+}$ atoms ($3d^4$) involves CI space to
diagonalize of about 60 billions configurations using the CAS+DDCI method and
of over 10 billions ones using the LCAS+S method. Among all those configurations, however, many of them are not
really important for the physics. Indeed, the major difference between the
case of metal atoms with a unique open shell and of metal atoms with multiple
open shells lies in the number of multiple-charge-transfer configurations. In
the former case, there are very few, while in the latter case there are the
most numerous configurations in the reference space, while their pertinence for
the low energy physics is far from obvious.
For instance, the $\rm Mn^{7+} Mn^{-}$ configurations exhibit a negligible
contribution in the wave functions of states involving two high-spin $\rm
Mn^{3+}$ ions, and contributes in a negligible way to the magnetic coupling
between the two ions.

From this simple illustration one sees that the configurations involved in the
usual methods are far too numerous compared to the really necessary ones for a
proper description of the low-lying states physics. We shall thus find another
criterion allowing to further select among the LCAS+S configurations the
really pertinent ones. Let us make a perturbative analysis of
equation~\ref{lcas}. The so-called zeroth-order configurations are the
configurations with large weight in the wave functions and the basis of any
minimal model. As the Hamiltonian only contains one and two particle
interactions, and using $|\Psi_m^0\rangle$ as zeroth-order wave functions, the
quasi-degenerate perturbation theory tells us that only the singly-excited and
doubly-excited configurations on $|\Psi_m^0\rangle$ are of real importance for
the physics.

The method we propose in this work for the evaluation of local interactions
takes advantage of the above analyses.  The reference part in
equation~\ref{lcas} shall be reduced to i) the zeroth-order part and ii) the
{\em single} charge transfer configurations only. The screening part thus
includes the single excitations on the zeroth-order part, plus the double
excitations built from a single excitation on top of a single charge transfer
excitation.  Compared to the LCAS+S we removed the multiple metal-to-metal and
ligand-to-metal charge transfer configurations from the reference part, and
the associated screening effects. Compared to the CAS+DDCI method (i) we
removed the multiple metal-to-metal charge transfer configurations and
associated screening configurations and (ii) we restricted the double
excitations to the screening excitations on the ligand-to-metal charge
transfers (as proved sufficient by the LCAS+S method~\cite{lcas+s}).
Consequently the number of configurations is dramatically reduced.  Going back
to the two manganese example, the present method (referred to as Selected
Active Space + Single-excitations or SAS+S) will involve only about $20\times
10^6$ configurations to diagonalize instead of $10\times 10^9$ for the LCAS+S
and $60\times 10^9$ for the CAS+DDCI method.

We tested the accuracy of the  proposed method on the magnetic coupling
of three compounds with respectively 2, 3 and 4 open shells per magnetic atoms
and 3, 1 and 2 dimensionality of the magnetic interactions~: $\rm KNiF_3$, $\rm
Ba_2CoS_3$ and $\rm YMnO_3$.  Since for $\rm Ba_2CoS_3$ and $\rm YMnO_3$ the
CAS+DDCI and LCAS+S methods are out of reach, we will present, for the sake of
comparison between the different methods, calculations where the treatment of
the screening effects are strongly pruned (only about 35\% of the screening
excitations are taken into account, see appendix for details). The evaluation
of the exchange integrals will thus not be physically correct, however we will
be able to compare the different methods. In a second time, we will use the
method proposed in this work (SAS+S) using fully screened calculations in
order to compare the computed values with experimental evaluations. Finally,
for the sake of analysis of the relative importance of the multiple
metal-to-metal and ligand-to-metal charge transfer configurations, we made an
intermediate calculation where only the multiple ligand-to-metal charge
transfer configurations were removed from the references. 
Since in each of the present examples the magnetic exchange are supposed to
follow and Heisenberg Hamiltonian, 
$$H=\sum_{<i,j>} J_{ij}\; \vec S_i \cdot \vec S_j  $$ 
we can extract the exchange integrals from
different spin excitations and, as a by product,  verify the validity of the
Heisenberg model.

\begin{table}
\begin{center}
\begin{tabular}{lccc}
\hline
\multicolumn{4}{c}{$\rm KNiF_3$ Calculations (meV)} \\
Calculation &  $\rm N_{CI}/10^6$ & $\rm J_{01}$ & $\rm J_{12}$ 
  \\ \hline
  CAS+DDCI &  28  &  6.89 & 7.06 \\
  LCAS+S &  1.1 &   6.71 & 6.82 \\
 Intermediate &  0.6 & 6.70 & 6.81 \\
 SAS+S  & 0.8 &  6.68 &  6.79 
 \\ \hline
\end{tabular} \hfill \\[2ex]

\begin{tabular}{lcccc}
\hline
\multicolumn{5}{c}{$\rm Ba_2CoS_3$ Pruned calculations (meV)} \\
Calculation &  $\rm N_{CI}/10^6$ & $\rm J_{01}$ & $\rm J_{12}$
& $\rm J_{23}$  \\ \hline
 CAS+DDCI &  34 & 1.86 & 1.85 & 1.84\\
 LCAS+S &  21 & 1.73 & 1.72 & 1.71 \\
 Intermediate &  4.4 & 1.72 & 1.71 & 1.70 \\
 SAS+S  &  2.3 & 1.62 & 1.62 & 1.61 \\ \hline
\end{tabular} \hfill \\[2ex]

\begin{tabular}{lcc}
\hline
\multicolumn{3}{c}{$\rm YMnO_3$ Pruned calculations (meV)} \\
Calculation & $\rm N_{CI}/10^6$ & $\rm J_{34}$  \\ \hline
 CAS+DDCI &  35 & 1.51 \\
 LCAS+S & 415 & - \\
 Intermediate &  5.7 & 1.43 \\
 SAS+S  &  1.5 & 1.36 \\ \hline
\end{tabular}

\caption{Comparison of different methods using a pruned evaluation of the
  screening effects. SAS+S refers to the method proposed in the present work 
  and intermediate
  refers to the calculation where only multiple ligand-to-metal charge
  transfer configurations were removed from the reference space. 
  $\rm N_{CI}$ is the size of the CI space. 
  $\rm J_{ij}$ correspond to the exchange integrals
  computed from the energy difference between the $S=i$ and $S=j$ spin states.}
\label{t:meth}
\end{center}
\end{table}

Table~\ref{t:meth} displays the evaluation of the magnetic exchange of the
three compounds for the two reference methods, the SAS+S method and the
intermediate calculation. One sees that for $\rm KNiF_3$ the four methods
yield very similar results with a relative difference of at most 3.5\%. For
$\rm Ba_2CoS_3$ and $\rm YMnO_3$ where the number of open shells is larger,
the error of the present method compared to the reference ones is somewhat
larger, of the order of 5 to 10\% according to the reference method.
Comparing now the LCAS+S and the intermediate calculations, one sees that the
difference is negligible (less than 1\%). The multiple ligand-to-metal
charge-transfer configurations, present in the LCAS+S calculation and not in
the intermediate one are thus of no influence on the exchange couplings, and
more specifically on the super-exchange part of them. The role of the multiple
metal-to-metal charge transfers can now be evaluated by the comparison between
the intermediate and SAS+S calculations. This time their influence is somewhat
larger for the $\rm Ba_2CoS_3$ and $\rm YMnO_3$ compounds, accounting for a
relative decrease of the super-exchange term of about 5\%.

\begin{largetable}
\begin{tabular}{lccccc}
\hline
Material & B3LYP & CAS+DDCI & LCAS+S & SAS+S & Exp. \\ \hline
$\rm KNiF_3$ & 14.86~\cite{Ni:b3lyp} & 6.98 & 6.77 & 6.74 & 7.7$\pm$0.4~\cite{KNF:exp} \\
$\rm Ba_2CoS_3$ & 4.08 &  - & - & 3.07 & 3.19$\pm$0.2~\cite{BCS:exp} \\
$\rm YMnO_3$ & 0.59 & - & -& 2.8 & 3~\cite{YMO:exp}
\\ \hline
\end{tabular}
\caption{Exchange integrals evaluation. Comparison with experiments (in meV).}
\label{t:exp}
\end{largetable}

Table~\ref{t:exp} displays the evaluation of the magnetic exchange for the
three examples using the present SAS+S method with full calculation of the
screening effects.  Comparing ab initio results with the experimental data, on
$\rm KNiF_3$, $\rm Ba_2CoS_3$ and $\rm YMnO_3$, the magnetic exchange
integrals are obtained within good accuracy to the experimental
values. Indeed, the error to the nominal value is respectively of 9\%, 4\% and
7\%, and within the error bar for $\rm Ba_2CoS_3$, that is of the same order
of magnitude of the best methods (CAS+DDCI and LCAS+S) for the single
open-shell systems such as vanadium or copper
oxides~\cite{dJCuO2}. We also like to point out that the
  experimental incertitudes should be taken with caution since there is not a
  real proper way to extract the magnetic coupling from susceptibility
  measurements, specifically in 2D and 3D systems. In this respect the
  experimental incertitudes are most likely larger than proposed on the 2D
  $\rm YMNO_3$ and 3D $\rm KNIF_3$ systems.

Let us sum up the main ideas supporting the present method. The CI
  space to diagonalize is built from three types of configurations. (i) The
  zeroth-order reference, that is the dominant configurations. These
  configurations are usually obvious from a simple analysis of the local
  physics. When the magnetic fragments are high-spin atoms they do correspond
  to the determinants for which the two magnetic atoms remain in a high-spin
  state.  (ii) The configurations dominant in the mediation of the magnetic
  coupling. In most cases it does correspond to simple ligand-to-metal and
  metal-to-metal charge transfer configurations. However, in some complex
  cases, as for instance in cases where ligands present low-lying excited
  states involving bridging orbitals (occupied and virtual), the dominant
  configurations in the magnetic coupling mediation also involve
  metal-to-ligand charge transfers and the associated the low-lying
  ligand-to-ligand excitations. (iii) The screening effects on all the above
  cited reference configurations, that is all single excitations on all above
  reference configurations. 
  Following these specifications our method is expected to be very general,
  applying to hetero-nuclear dimers of high-spin atoms, to complex ligands
  (extended, non-symmetric, conjugated, etc.), even to extended magnetic
  centers. 
 
  Let us also point out that the present method, that addresses the problem of
  numerous open-shells magnetic centers, can be profitably combined with
  complementary techniques developed for the reduction of the CI space as for
  instance~: the treatment of large ligands using localized
  orbitals~\cite{Barone} or a more efficient treatment of the screening
  effects by the optimization of the virtual orbitals~\cite{virtuelles}. 

In conclusion, the SAS+S method presented in this work aim at the evaluation
of magnetic couplings between atoms with multiple open shells.  Indeed, on one
hand, density functional theory encounters a lot of difficulties in the
description of such strongly correlated systems. These difficulties are
particularly dramatic for the evaluation of magnetic couplings since it often
fails even to get the correct value by a factor two or more. On the other
hand, embedded fragment quantum chemical ab initio methods succeeded in
evaluating magnetic couplings when the magnetic atoms display only one or two
open-shells. For larger numbers of open-shells, the size of the calculations
were out of reach. We proposed in this paper a new CI method that condenses the
pertinent information on magnetic excitations.  This method allowed us to
reach accurate ab initio evaluation of effective magnetic couplings in high
spin cobalt or manganese compounds. It is the first time that such couplings
become reachable. The SAS+S method thus opens new fields of research. For
instance one can think to look at the local fluctuations of magnetic couplings
in manganites, fluctuations known to be important for the colossal
magneto-resistance effects but not experimentally directly accessible.

\acknowledgments The authors thank Daniel Maynau for providing them with
the CASDI suite of programs. These calculations where done using the
CNRS IDRIS computational facilities under project n$^\circ$1842 and
the CRIHAN computational facilities under project n$^\circ$2007013. The present
project was financed by the french ANR project SEMOME. 

\bigskip 

 \newcounter{section}
 \newcounter{subsection}

\begin{figure}[h]
\begin{center}
\resizebox{6cm}{!}{\includegraphics{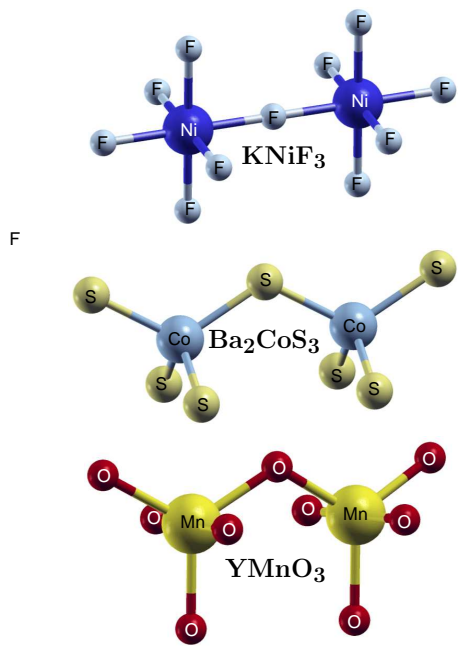}}
\end{center}
\caption{Quantum part of the embedded fragments.}
\label{fig:frag}
\end{figure}

\appendix{\bf Computational details} 
The embedded fragments used for the magnetic couplings were built as follow.
i) A quantum part containing the two magnetic atoms, the ligands mediating the
interactions between them (S,F or O atoms) and their first shell of
coordination (see figure~\ref{fig:frag}).  ii) The embedding part reproducing the main effects of the rest
of the crystal on the quantum part~: the exclusion effects and the Madelung
potential. The exclusion effects are treated using two shells of total ions
pseudo-potential~\cite{TIPS} that forbid to the quantum part electrons to
delocalize out of the fragment. The Madelung potential is computed using an
appropriate renormalized set of charges~\cite{env2} with an accuracy better
than $10^{-2}$meV. 

The basis sets and pseudo-potentials used in this work are of valence $3\zeta$
quality on the metal atoms and $3\zeta+P$ on the ligands. They can be found in
reference~\cite{bases}.

The fragment orbitals were optimized using a Complete Active Space Self
Consistent Field calculation on the highest spin state. The active orbitals
were chosen as the magnetic opened $3d$ orbitals. This method is well known to
yield a good separation between the occupied, active and virtual orbitals used
in our reference spaces. The ligand orbitals mediating the
  interactions (used in the LCAS+S and SAS+S methods) were extracted from the
  occupied orbitals in a two-steps process. First, the ligand orbitals which
  occupation number fluctuates between different states (computed at a the
  minimal CAS+S level) were extracted from the Fermi sea using a
  multi-state natural orbitals method~\cite{msnatorb}.  Second, among these
  orbitals the bridging ones were obtained by maximizing the hopping (kinetic
  integral) between them and the magnetic orbitals.  The bridging orbitals
  used in the present work are thus uniquely defined and qualitatively
  correspond to the bridging orbitals expected from simple physical 
  considerations\cite{BCS:exp}. 



\begin{thebibliography}{9999}


\bibitem{Revue} M.B. Lepetit, {\em
    Recent Research Developments in Quantum Chemistry 3}, p. 143, Transword
  Research Network (2002).


\bibitem{env} M.B. Lepetit, N. Suaud, A. Gell\'e and V. Robert,
  J. Chem. Phys. {\bf 118}, 3966 (2003)~;  A. Gell\'e and M.B. Lepetit,   
J. Chem. Phys. {\bf 128}, 244716 (2008). 

\bibitem{DDCI} J. Miralles, J. P. Daudey and R. Caballol,
  Chem. Phys. Lett. {\bf 198}, 555 (1992)~; V. M. Garc\'ia {\it et al.},
  Chem. Phys. Lett. {\bf 238}, 222 (1995)~; V. M. Garc\'ia, M. Reguero and
  R. Caballol, Theor. Chem. Acc. {\bf 98}, 50 (1997).

\bibitem{lcas+s} A. Gell\'e, M. L. Munzarov\'a, M.B. Lepetit and F. Illas,
  Phys. Rev. {\bf B 68}, 125103 (2003)~; C J. Calzado, J. F. Sanz and
  J. P. Malrieu, J. of Chem. Phys. {\bf 112}, 5158 (2002).


\bibitem{CALZADO99} See for instance~: C. Jimenez Calzado, J. Fernandez Sanz,
  J. P. Malrieu and F. Illas, {\it Chem. Phys. Lett.} {\bf 307}, 102 (1999);
  D. Munoz, F. Illas, I. de P.R. Moreira, {\it Phys. Rev. Letters} {\bf 84},
  1579 (2000); A. Gell\'e and M.B. Lepetit, Phys. Rev. {\bf B 74}, 235115
  (2006). 

\bibitem{david} D. Munoz ,C. De Graaf ,F. Illas, J. of Comp. Chem. {\bf 25},
  1234 (2004).


\bibitem{molmagn} J. Cabrero {\it et al.}, J. Phys. Chem. {\bf A 104}, 
9983 (2000). 

\bibitem{nav2o5}
  N. Suaud, M.B. Lepetit, Phys. Rev. Lett. {\bf 88}, 056405 (2002). 


\bibitem{dJCuO2}
I. de P. R. Moreira and F. Illas, Phys. Rev. {\bf B 55}, 4129 (1997).


\bibitem{Ni:b3lyp} X. Feng and N. M. Harrison, {\it Phys. Rev.} {B 70}, 092402
  (2004).

\bibitem{KNF:exp} M. E. Lines, Phys. Rev. {\bf 164}, 736 (1967).

\bibitem{BCS:exp} A. D. J. Barnes, T. Baikie, V. Hardy, 
M.B. Lepetit, A. Maignan, N. A. Young and M. Grazia
Francesconi, J. Mater. Chem. {\bf 16}, 3489 (2006).

\bibitem{YMO:exp} J. Park, J.-G. Park, G. S. Jeon, H.-Y. Choi, C. Lee, W. Jo,
  R. Bewley, K. A. McEwen, and T. G. Perring, Phys. Rev. {\bf B 68}, 104426
  (2003).


\bibitem{Barone} C. Angeli, C. J. Calzado, R. Cimiraglia, S. Evangelisti,
  N. Guihery, Th. Leininger, J. P. Malrieu, D. Maynau, J. V. Pitarch Ruiz and
  M. Sparta, {\it Mol. Phys.} {\bf 101}, 1389 (2003)~; V. Barone, I. Cacelli
  and A. Ferretti, {\it J. Chem. Phys.} {130}, 094306 (2009).

\bibitem{virtuelles}  V. Barone, I. Cacelli, A. Ferretti, and G. Prampolini,
  {\it Phys. Chem. Chem. Phys.} {\bf 11}, 3854 (2009). 

\bibitem{TIPS} N. W. Winter, R. M. Pitzer and D. K. Temple, {\it
J. Chem. Phys.} {\bf 86}, 3549 (1987).


\bibitem{env2} A. Gell\'e and M.B. Lepetit, J. Chem. Phys. {\bf 128}, 244716
  (2008).


%
%
%
%
\bibitem{bases} Z. Barandiaran and L. Seijo, Can. J. Chem. {\bf 70}, 409
  (1992).

\bibitem{msnatorb} A. Gell\'e and M.B. Lepetit,  unpublished.\\
  The multi-state natural orbital method is an adaptation of the joint
  diagonalization method to the determination of natural orbitals
  diagonalizing simultaneously the density matrix of several states. 

\end{thebibliography}
\end{document}